\documentclass[twocolumn,amsmath,amssymb,aps,prb]{revtex4}

\usepackage{graphicx}
\usepackage{dcolumn}
\usepackage{bm}

\begin{document}

\preprint{PREPRINT}
\title{Effects of non-universal large scales on conditional structure
functions in turbulence}

%

\author{Daniel B. Blum}\
\author{Surendra Kunwar}
\author{James Johnson}
\author{Greg A. Voth}\

\affiliation{Department of Physics, Wesleyan University, Middletown,
CT  06459, U.S.A.}

\homepage{http://gvoth.web.wesleyan.edu/lab.htm}

\date{\today}

\begin{abstract}

We report measurements of conditional Eulerian and Lagrangian
structure functions in order to assess the  effects of non-universal
properties of the large scales on the small scales in turbulence. We
study a 1m $\times$ 1m $\times$ 1.5m flow between oscillating grids
which produces $R_\lambda=285$ while containing regions of nearly
homogeneous and highly inhomogeneous turbulence. Large data sets of
three-dimensional tracer particle velocities have been collected
using stereoscopic high speed cameras with real-time image
compression technology.  Eulerian and Lagrangian structure functions
are measured in both homogeneous and inhomogeneous regions of the
flow. We condition the structure functions on the instantaneous
large scale velocity or on the grid phase. At all scales, the
structure functions depend strongly on the large scale velocity, but
are independent of the grid phase. We see clear signatures of
inhomogeneity near the oscillating grids, but even in the
homogeneous region in the center we see a surprisingly strong
dependence on the large scale velocity that remains at all scales.
Previous work has shown that similar correlations extend to very
high Reynolds numbers. Comprehensive measurements of these effects
in a laboratory flow provide a powerful tool for assessing the
effects of shear, inhomogeneity and intermittency of the large
scales on the small scales in turbulence.

\end{abstract}

\maketitle

\section{\label{sec:intro}Introduction}

Many of the most powerful insights in the study of fluid turbulence
are rooted in the idea of an energy cascade where the chaotic
process of transferring energy to smaller scales allows the small
scales to become universal and independent of the details of the
forcing mechanism at the large scales. However, careful examination
of many small scale statistics in different flows has
shown that the reality of turbulence is quite a bit more             
complicated. Some statistics such as the scaling exponents of
Eulerian and Lagrangian structure functions are nearly identical in
different flows~\cite{Sreenivasan:1997:PST, arneodo:1996,
Biferale:2008}. But other small scale statistics such as the
coefficients in scaling laws~\cite{Ouellette:2006,biferale:2001} or
the scalar derivative skewness~\cite{warhaft:2000} show dependence
on the properties of the large scales up to the largest Reynolds
numbers measured.

A traditional approach to deal with dependence on the large scales
has been to classify flows, and allow that there might be
differences between categories of flows such as free shear flows
(jets, mixing layers, \textit{etc}), wall bounded shear flows
(boundary layers, channel flows, \textit{etc}), or isotropic
turbulence (wind tunnel grid turbulence or numerical simulations in
a box with periodic boundary conditions, \textit{etc}).  A careful
empirical comparison of statistics between different flows could
then show which properties of the small scales are truly independent
of the large scales.


Recent developments in experimental tools have allowed for the
measurement of small scale Lagrangian statistics~\cite{toschi:2009}.
In contrast to flows designed for Eulerian measurements such as
hot-wire anemometry, Lagrangian measurements do not require large
mean velocity in order to use Taylor's frozen flow hypothesis.  In
fact the opposite is desired since small mean velocity allows a
particle to be tracked in an observation volume for the longest
possible time.  This has led to significantly different flow designs
for Lagrangian measurements.  Many new flows have been introduced
which have complex large scales and are difficult to place into
traditional categories which were originally meant for flows
designed for hot-wire anemometry.  Two widely used examples of flows
with low mean velocity are counter rotating disks~\cite{Douady:1991,
Voth:2002, Mordant:2001} and oscillating
grids~\cite{Fernando:1993,Fernando:1994, Ott:2000}.  A yet newer
generation of flows are currently under study including the random
jet array~\cite{variano:2008}, corner stirred
tank~\cite{Luthi:2007}, Lagrangian exploration module (LEM), and
radial acoustic jets which each have unique large scales.

Initial work on Lagrangian measurements in flows with complex large
scales and a small mean velocity has primarily assumed that the
small scale statistics of interest are independent of the large
scale forcing of the flow. This assumption has been tested in
several cases by thorough quantitative comparison of small scale
statistics in different flows. Comparison of acceleration
probability distribution functions (pdf) in direct numerical simulations
(DNS) and counter
rotating disk experiments are found to be flow independent ~\cite{Voth:2002,Yeung:1999,Bodenschatz:2004}                          
.  In addition, the scaling exponents of the Lagrangian
 structure functions have been compared between DNS
and experiment, and found to be in close
agreement~\cite{Biferale:2008}.  Much more work is needed to
determine the degree to which the large scales of different flows
affect various Lagrangian statistics.

However, there is a more direct way to evaluate whether the small
scales in a flow are independent of the large scales: the small
scale measurements can be conditioned on a measurement of the state
of the large scales.  Two previous studies have shown strong
dependence of the small scales of Eulerian structure functions on
the instantaneous velocity, which is dominated by the large scales.
Praskovsky \textit{et al.}~\cite{Praskovsky:1993} extensively study
interactions between the large scales and inertial range scales in
two high Reynolds number wind tunnel flows.  Strong correlations
between the large scales and the velocity structure functions are
found at all length scales. They interpret this as being consistent
with the correct application of Kolmogorov theory with a fluctuating
energy injection at the large scales. Sreenivasan and
Dhruva~\cite{sreenivasan:1998} measured Eulerian velocity structure
functions from atmospheric boundary layer data for $R_\lambda >
10^4$, some of the largest Reynolds numbers ever measured.  They
find that the structure functions conditioned on the large scale
velocity show a strong dependence, and they show that
 DNS and grid turbulence measurements show almost no dependence. They
attribute the dependence to large scale shear, and show how to
remove the effect to improve power law scaling.
   There is evidence that other small scale
statistics show conditional dependence on the large scales.  The acceleration variance shows a
strong dependence when it is conditioned on the large scale
velocity~\cite{voth:2000,Sawford:2003}.

One challenge in discussing interactions between large scales and small scales is the
 very non-universal nature of the large scales.  Each flow has a
unique set of large scales, which may depend on time, geometry, or
driving parameters. So it has been difficult to isolate the aspects
of the large scale flow that are affecting the small scales.
Anisotropy is the aspect that is best understood. Extensive work has
identified persistent anisotropy at small scales even at very high
Reynolds numbers~\cite{Sreenivasan:1991:LIP, warhaft:2000}, and
analysis using spherical tensor decomposition has placed this
problem on solid footing ~\cite{arad:1999,biferale:1999}. However,
this is not the only effect of the large scales. Here we wish to
distinguish two additional aspects of the large scales that are
particularly important. Inhomogeneity is the spatial variation of
statistics. Large scale intermittency is temporal fluctuations on
time scales longer than the eddy turnover
time, $L/u$.  Both inhomogeneity and large scale intermittency often occur together in real flows, but are         
distinct properties since flows can be conceived that have each
without the other.  For example, a homogeneous turbulent flow in DNS
can have large scale intermittency by having the energy injection
varied in time.                                                    

In this paper we present a comprehensive set of measurements of the
dependence of Eulerian and Lagrangian velocity structure functions
conditioned on the large scale velocity.  We use a flow between two
oscillating grids which is relatively homogeneous in the central
region, but has large inhomogeneity near the grids.     This allows
us to isolate the signatures of different properties of the large
scales. We find clear signatures of inhomogeneity, but a significant
part of the dependence of the structure functions on the large scale
velocity seems to be the result of large scale intermittency.  A
better understanding of this dependence on non-universal large
scales will help in the identification of universal statistics, and
the comparison of different flows.

\section{\label{sec:design}Experiment}

\begin{figure}[tbh]
\begin{center}
\includegraphics[bb=0 0 640 480,scale=0.4]{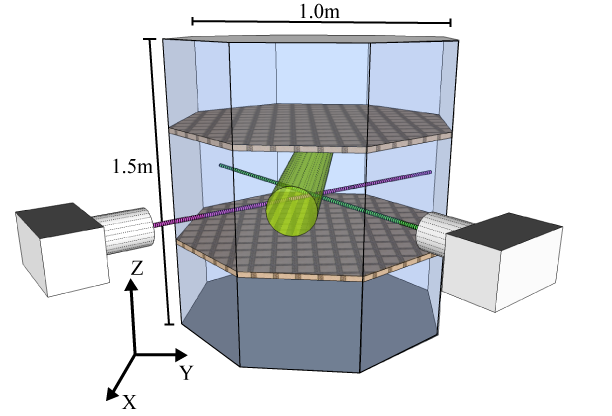}
\caption{\label{fig:tankdiagram} Experimental apparatus diagram. Two
oscillating grids were held 56.2 cm apart in a 1,100 liter octagonal
prism Plexiglas tank.  Two high speed cameras are used to
stereoscopically image chosen regions of the tank in order to record
3D particle positions. Illumination is provided by a Nd:YAG laser
with 50 W average power.}
 \end{center}
 \end{figure}

This work is based on optically tracking passive tracer particles
seeded in a turbulent flow agitated by two oscillating grids as shown in Fig.~\ref{fig:tankdiagram}.   For
clear measurements of 1-D inhomogeneity a large system is needed to
create significant separation for locally homogeneous and
inhomogeneous regions, and to create sufficiently high Reynolds
numbers.  In order to study large scale effects conditional
statistics were analyzed, which required large data sets ($>10^9$
particle pairs). Storage, speed, and budget concerns led to the
development of real time image compression
circuits~\cite{chan:2007}. These devices enabled nearly endless data
acquisition for a nominal cost.

\subsection{Experimental Apparatus}
 Turbulence was generated between two
identical octagonal grids oscillating in phase in an octagonal
Plexiglas tank that is  1m$~\times$ 1m$~\times$ 1.5m and filled with
approximately 1,100 liters (300 gallons) of filtered, degassed water. The grids
have 8 cm mesh size, 36\% solidity, and were evenly spaced from the
top and bottom of the tank with a 56.2 cm spacing between grids and
a 1 cm gap between grid and tank walls. The stroke was 12 cm peak to
peak, powered by an 11 kW motor. A typical grid frequency was 3 Hz,
but was raised to 5 Hz to investigate Reynolds number dependence.
Water cooling maintains the temperature at $\pm$ 0.1C during each
run. 
Neutrally buoyant 136 $\mu$m diameter polystyrene tracer particles
were added to the flow until approximately 50 were seen by each
camera. This particle density was chosen to maximize data per frame
while minimizing tracking errors. Particle density could greatly
increase with the planned addition of two more
cameras~\cite{ott:1999,Dracos:1996:TDV}. One difficulty of
oscillating grid flows is that vibrations from the oscillatory drive
can couple to the camera supports and degrade imaging accuracy.  We
mounted a custom camera support on an optical table to minimize
vibrations. Air bubble suppression was an additional concern. We
developed a method to keep all water seals and bearings sufficiently
wet to maintain an air tight seal.

\subsection{Detection}
These data were acquired using 3D PTV (particle tracking
velocimetry) measurements using two Bassler A504K video cameras
capable of 1280 $~\times$ 1024 pixel resolution at 500 frames per
second (a data rate of approximately 625 MB per second per camera).
Recording such a high data rate is a significant technological
hurdle. A typical system would store data in 4 GB of video RAM, so
that one run would last just 7 seconds before waiting approximately
7 minutes for the data to download to hard disk. We use an image
compression circuit to threshold images in real-time so that only
pixels above a user defined brightness limit are regarded as
particle data and retained while the dark background pixels are
discarded~\cite{chan:2007}. This technique produces a dynamic data
compression factor of 100-1000, which enables continuous data
collection and storage to hard disk.

Our first implementation of the image compression circuits have
faced two major challenges. First, the simple thresholding
compression reduces particle center accuracy. However, particle
finding accuracy is typically degraded by only 0.1 pixel, which is
typically less than the uncertainty in particle finding from
unthresholded images.  Secondly, because frame number information
was created and recorded separately on each computer, any operating
system delay can lead to frames lost and timing mismatch between the
cameras.  For the measurements in this paper, frame number errors
were corrected in post
processing.  Updated versions of the image compression circuit have solved this problem by including              
camera frame number in the data stream the computers record.

Particles are illuminated using a 532 nm pulsed Nd:YAG laser with 50
W average power. The beam was expanded to create an illumination
volume approximately  7cm $\times$ 4cm $\times$ 5cm.  Images were
processed to find the center of each particle as seen by each camera
and then stereomatched to find the 3D position in real space.
Stereomatching was accurate to approximately 11~$\mu$m (0.08
particle diameters or 0.2 pixels). At this level of accuracy it is
essential to have a very good calibration of camera position
parameters to use for stereomatching. We start with a traditional
calibration to obtain initial 3D stereomatching ~\cite{Otto:2008}.
We then use known stereomatched pairs from the two cameras, and run
a non-linear optimization to minimize the stereomatching error and
find optimal camera position parameters~\cite{ott:1999}.               

\section{Results}

\subsection{Characterizing the Flow}
We define a characteristic velocity by $u=(\langle u_i u_i\rangle
/3)^{1/2}$ and a characteristic length scale by $L= u^3/\varepsilon$
where $\varepsilon$ is the energy dissipation rate per unit mass
defined in section ~\ref{sec:E_disp}.  For the center region $u =
6.0$~cm/s, $L = 9.0$~cm, and for the near grid region $u =
8.3$~cm/s, $L = 4.5$~cm. The Taylor Reynolds number, $R_\lambda=(15
u L/\nu)^{1/2}$, (where $\nu$ is the kinematic viscosity) ranges
from 285 for 3 Hz grid frequency to 380 for 5 Hz grid frequency in
the center. Near the grid at 3Hz  $R_\lambda = 230.$  The Kolmogorov
length and time scales are $\eta= 140 \mu$m, $\tau_\eta=20$ ms in
the center region, and $\eta= 94~\mu$m, $\tau_\eta=8.8$ ms in the
near grid region.

\begin{figure}[tbh]
\begin{center}
\includegraphics[bb=0 0 640 480,scale=0.5]{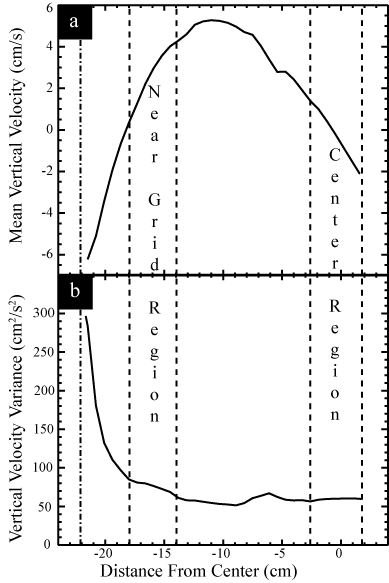}
\caption{\label{fig:meanvar_z} Mean and variance of vertical
velocity along the central vertical axis of the tank.  Grid
frequency 3 Hz, grid separation distance 56.2 cm.  The dot-dash line
represents the grid height at maximum amplitude.   The remainder of
the paper will focus on measurements in two
 regions designated by the vertical dashed lines:  one at the center of the tank and one near the grid.
}
 \end{center}
 \end{figure}

\begin{figure}[tbh]
\begin{center}
\includegraphics[bb=0 0 640 480,scale=0.4]{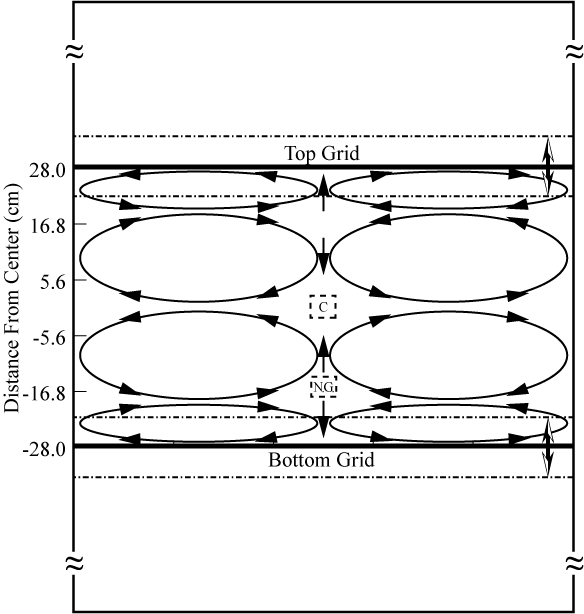}
\caption{\label{fig:flowdiagram} Scale diagram of 56cm $\times$
100cm area between grids showing the mean circulation torii which
are nearly rotationally symmetric about the central vertical axis.
Center and Near Grid observation volumes are drawn in dashed lines,
which shows the relative size and position of the observation
volumes in figure~\ref{fig:meanvar_z}.  Horizontal dot-dashed lines
represent range of motion of the top and bottom grids.}
 \end{center}
 \end{figure}

Figure{~\ref{fig:meanvar_z}(a)} shows the mean vertical velocity as
a function of the vertical position along the central axis of the
tank. The top and bottom grids are separated by 56.2 cm,
approximately 7 \emph{L}. In Fig.~\ref{fig:meanvar_z}, the
dot-dashed line indicates the maximum amplitude of the bottom grid,
22.1 cm below the center of the tank. Data was collected at 5
separate heights in order to measure the complete flow profile from
the center of the tank to the bottom grid. Mapping the bottom half
of the tank is sufficient because the geometrical symmetry produces
a mirror image above the midplane. The two volumes which we will
focus on throughout this paper are bounded by the dashed lines, and
will be referred to as the center (C), and near
grid (NG) observation volumes. At this grid separation distance the mean    
flow traces four torii, two above and two below the center plane of
the tank, as shown in the sketch in figure~\ref{fig:flowdiagram}
(drawn to scale). In the large central region, the effect of the
mean flow is to pump highly energetic fluid from the region near the
grid towards the center of the tank. In
figure{~\ref{fig:meanvar_z}(a)}, there are two points where the mean
vertical velocity reaches zero: one near the center, the other 18 cm
from the center just below the near grid observation volume. The
existence of this second stagnation point and reverse circulation
region depicted in Fig.~\ref{fig:flowdiagram} is a common feature in
mean flows generated by oscillations~\cite{Otto:2008}. In the
following measurements, the small mean velocity has been subtracted
so that we study the fluctuating velocity field.

Figure{~\ref{fig:meanvar_z}(b)} shows the vertical velocity variance
along the central axis as a function of vertical position. The
velocity variance is large near the grid and quickly falls off
towards the center where it is nearly homogeneous. The center and
near grid observation volumes were chosen to provide a contrast
between the large homogeneous region in the center and the much more
inhomogeneous region near the grid.  In the center, the variance of
the velocity is homogeneous for several \emph{L} in either
direction.  The velocity variance ranges moderately in the near grid
observation volume, and enormously within one \emph{L} below this
region. In Fig.~\ref{fig:meanvar_z}, deviations from a smooth curve
are not due to statistical uncertainty, but are a result of patching
5 calibrated regions together with the majority of error coming from
measuring absolute position in the tank.

It is interesting to note that we made measurements in a flow with a
smaller grid separation of 35 cm and found that the Reynolds number
in the center was lower. The characteristic velocity in the center
did increase due to the closer proximity of the grids, but $L$ was
reduced by a larger amount resulting in approximately 8\% decrease
in $R_\lambda$.  The reason for the unexpected decrease in Reynolds
number is a reversal of the mean velocity compared with larger grid
separations. For larger grid separation distances, energetic fluid
from near the grids is carried to the center by the mean flow.
However, at 35 cm grid separation the mean velocity reverses which
results in a lower Reynolds number in the center.

\subsection{Structure Functions}
\begin{figure}[tbh]
\begin{center}
\includegraphics[bb=0 0 640 480,scale=0.6]{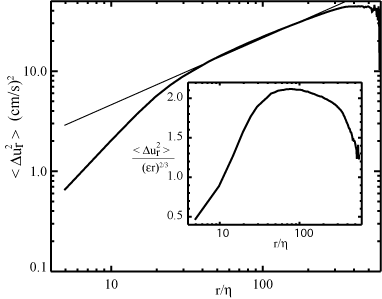}
\caption{\label{fig:sf2} Eulerian second order longitudinal velocity
structure function shown as a function of pair separation $r$ scaled
by the Kolmogorov length $\eta$. The inset shows this data
compensated by Eq.~\ref{1} for $p$=2.}
 \end{center}
 \end{figure}

\begin{figure}[tbh]
\begin{center}
\includegraphics[bb=0 0 640 480,scale=0.6]{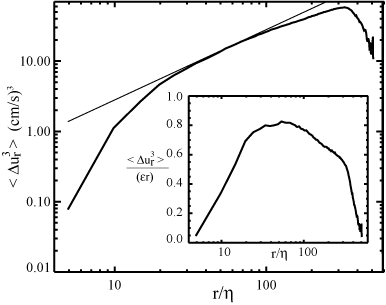}
\caption{\label{fig:sf3} Eulerian third order longitudinal velocity
structure function. The inset shows this data compensated by
 Eq.~\ref{1} for $p$=3.}
 \end{center}
 \end{figure}

To measure Eulerian structure functions we first find the instantaneous               
longitudinal velocity difference between two particles a distance
$r$ apart $\Delta
u_r=[\mathbf{u}(\mathbf{x})-\mathbf{u}(\mathbf{x}+\mathbf{r})]_L$,
where the $L$ subscript denotes the longitudinal component, found by
projecting the 3D velocity difference vector onto the vector
connecting the two particles.  The                  
longitudinal structure functions are defined as $D_p=\langle (\Delta
u_r)^p \rangle$ where $p$ represents the order of the structure
function and the brackets represent the ensemble average. In the
inertial range, Kolmogorov (1941) gives
\begin{equation}
 \langle \Delta u_r^p \rangle=
C^{(E)}_p(\varepsilon r)^{p/3}, \label{1}
\end{equation}
where the $C^{(E)}_p$ are Eulerian Kolmogorov constants and
$\varepsilon$ is the energy dissipation rate.

Figures~\ref{fig:sf2} and~\ref{fig:sf3} show the measured second and
third order longitudinal velocity structure functions with the
straight thin lines representing Kolmogorov's prediction from
Eq.~\ref{1}. The insets show the structure functions compensated by
Eq.~\ref{1}. At $R_\lambda=285$, any scaling range is very limited,
but the plateaus can be used to estimate the inertial range.


Lagrangian structure functions were measured from temporal velocity
differences along a particle trajectory.  The velocity difference
now becomes $\Delta u_\tau=u(t)-u(t+\tau)$, where $\tau$ is the time
interval between measurements.  We use the vertical component of the
velocity for Lagrangian velocity differences throughout this paper,
although results for the other components are similar.  For
Lagrangian structure functions Kolmogorov (1941) predicts
\begin{equation}
 \langle \Delta u_\tau^p \rangle =
C^{(L)}_p(\varepsilon \tau)^{p/2}. \label{2}
\end{equation}

\subsection{Energy Dissipation Rate Measurement}
\label{sec:E_disp} The energy dissipation rate $\varepsilon$ is an
important value throughout this analysis, it is worth a moment to
discuss how it is determined.  Limitations in particle density
preclude direct measurement via the definition
$$\varepsilon = 2\nu \langle s_{ij} s_{ij}\rangle$$
with
$$s_{ij} = \frac{1}{2}\left(\frac{\partial u_i}{\partial x_j}+\frac{\partial u_j}{\partial x_i}\right).$$
Instead we utilize Kolmogorov's 4/5 law:  Eq.~\ref{1} with $p=3$
where the coefficient $C^{(E)}_3=-4/5$ can  be derived from the
Navier-Stokes equations. We identify the inertial range with the
plateau in the compensated third order structure function
(Fig.~\ref{fig:sf3} inset).  The inertial range is chosen to be 25
to 91 $r/\eta$ (.35 to 1.3 cm). If the same inertial range is used
in the second order structure function, the energy dissipation rate
determined from it (using the empirical coefficient
$C^{(E)}_2=2.0$)~\cite{Pope:2000} is within 3\% of the value
calculated from the third order.

\subsection{Phase Dependence}
A simple energy cascade has constant energy input at the largest
length scales. An obvious departure from constant energy input is
the oscillating grid driving mechanism. The sinusoidal motion of the
grid directly corresponds to energy input with periodic time
dependence. It seems likely that such a strongly periodic energy
input would have a signature throughout the whole energy cascade.

The method we employ throughout this work to detect signatures of
the large scales is to condition various statistics on some
measurement of the state of the large scales.  In this case, we
condition on the phase of the grid motion, $\phi$.  Conditioning
instantaneous single particle statistics such as the mean and
variance of the velocity shows some sinusoidal dependence on grid
phase. For example, in the center the conditional variance, $\langle
(u-\langle u \rangle)^2 | \phi \rangle$, varies by $1\%$ over the
cycle of the grid.  In the near grid region, the conditional
variance varies by $10\%$. The mean vertical velocity in the center,
$\langle u | \phi \rangle$, varies by 0.8 cm/s over a cycle of the
grid which is 10\% of the standard
deviation. Near the grid, the conditional mean velocity varies by 2 cm/s, which is 20\% of the             
standard deviation at that location.

\begin{figure}[tbh]
\begin{center}
\includegraphics[bb=0 0 640 750,scale=0.5]{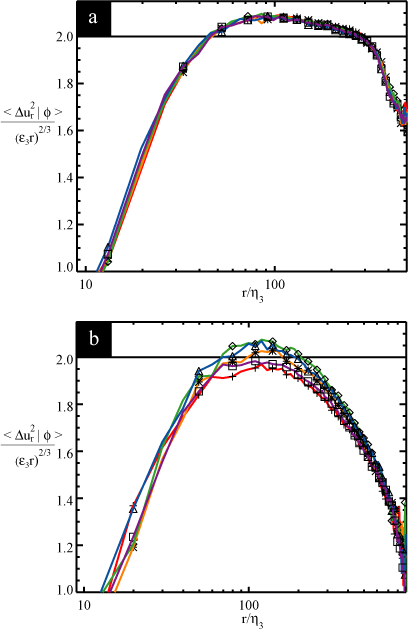}
\caption{\label{fig:sf2_comp_cond_phase} (Color Online) Second order
compensated velocity structure functions conditioned on grid phase.
The collapse shows the very weak phase dependence.  a) center of the
tank~(C).  b) near the grid~(NG). Zero and 2$\pi$ phase represents
grid at lowest possible amplitude. $\phi$: $+$ = 0 - 2$\pi$/5,
$\ast$ = 2$\pi$/5 - 4$\pi$/5, $\diamond$ = 4$\pi$/5 - 6$\pi$/5,
$\triangle$ = 6$\pi$/5 - 8$\pi$/5 , $\Box$ = 8$\pi$/5 - 2$\pi$.}
 \end{center}
 \end{figure}

Figure{~\ref{fig:sf2_comp_cond_phase}} shows the compensated second
order longitudinal structure functions conditioned on the phase,
$\langle \Delta u_r^2 |\phi \rangle$.  In the center of the flow
(Fig.{~\ref{fig:sf2_comp_cond_phase}}(a)) the structure functions
have essentially no change with phase.  Near the grid
(Fig.{~\ref{fig:sf2_comp_cond_phase}}(b)) there is a slight
dependence on grid phase. To emphasize the differences between
structure functions at different phases, we compensated the
structure functions by a single energy dissipation rate in each
figure, $\varepsilon_3=24.6$ cm$^2/$s$^3$ in the center and
$\varepsilon_3=131 $ cm$^2/$s$^3$ near the grid. These values were
determined when the grid is in mid amplitude (the third bin). The
good collapse of the structure functions at all phases across the
entire range of $r$ shows the minimal dependence of the small scales
on the large scale periodicity of the flow created by the
oscillating grids.  One possible source of dependence of small scale
statistics on large scales has been shown to be minimal.

\subsection{Dependence on Large Scale Velocity}
\label{sec:Large Scale Dependence}
\subsubsection{Eulerian structure functions conditioned on the large
scale velocity:  center region}

A more revealing dependence on the large scales of the flow is found
by conditioning the velocity structure functions directly on the
large scale velocity. A convenient measurable quantity that reflects
the local instantaneous state of the large scales is the average
velocity of the particle pair used for the structure function,
defined as $\Sigma u_z= (u_z(\mathbf{x}) +
u_z(\mathbf{x}+\mathbf{r}))/2.$ Alternatively, conditioning on the
average velocity of many particles, not just one pair, was studied
and found to have very similar results, but we choose to focus on
$\Sigma u_z$ because it can be more easily measured and does not
depend on the observation volume and seeding density.  Additional
conditioning quantities will be discussed in
section~\ref{sec:Discussion}.

\begin{figure}[tbh]
\begin{center}
\includegraphics[bb=0 0 640 750,scale=0.5]{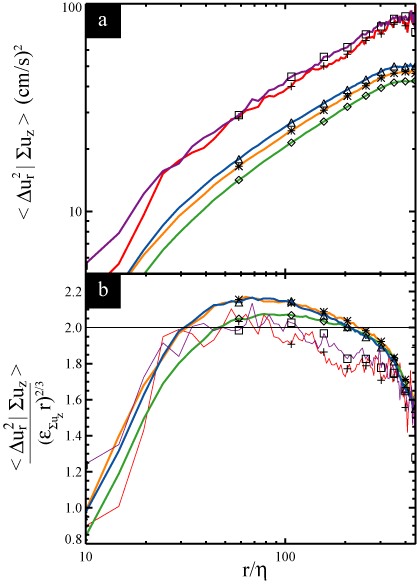}
\caption{\label{fig:sf2_zvel_center} (Color online) Second order
velocity structure function conditioned on particle pair velocity
(vertical component) in the center of the tank.  a) Uncompensated
structure functions b) Individually compensated by the energy
dissipation rate for each conditional data set. Symbols represent
the following dimensionless vertical velocities, $\Sigma
u_z/\sqrt{\langle u_z^2 \rangle}$: $+$ = 4.2 to 2.5, $\ast$ = 2.5 to
0.84, $\diamond$ = 0.84 to -0.84, $\triangle$ = -0.84 to -2.5,
$\Box$ = -2.5 to -4.2.}
 \end{center}
 \end{figure}

Figure{~\ref{fig:sf2_zvel_center}(a)} shows the second order
Eulerian velocity structure function conditioned on $\Sigma u_z$.   The
smallest values of the structure function correspond to pair
velocities near zero, represented by $\diamond$, while large $|\Sigma u_z|$
results in larger values of the structure functions.   For the bins
we chose, the structure function conditioned on large values of
$\Sigma u_z$ is nearly twice the value when conditioned on $\Sigma u_z$ near zero.

Figure{~\ref{fig:sf2_zvel_center}(b)} shows the data in
Figure{~\ref{fig:sf2_zvel_center}(a)} compensated by Kolmogorov
inertial range scaling.  The functional forms are quite similar,
confirming the impression from Fig.{~\ref{fig:sf2_zvel_center}(a)}
that all length scales are affected similarly by the instantaneous
state of the large scales.  In Fig.{~\ref{fig:sf2_zvel_center}(b)}
we used a different energy dissipation rate, $\varepsilon_{u_z}$ to
compensate each of the 5 individual large scale vertical velocity
bins. This insures all conditions plateau at approximately the same
value, and allows for direct comparison of the functional forms of
the conditional structure functions.

The strong dependence of the conditional structure functions on the
large scale velocity at all scales reveals that the small scales are
not statistically independent of the large scales in this flow.
There is not any detectable trend toward the smaller scales becoming
less dependent on the large scale velocity than somewhat larger
scales.

 \begin{figure}[tbh]
\begin{center}
\includegraphics[bb=0 0 640 480,scale=0.5]{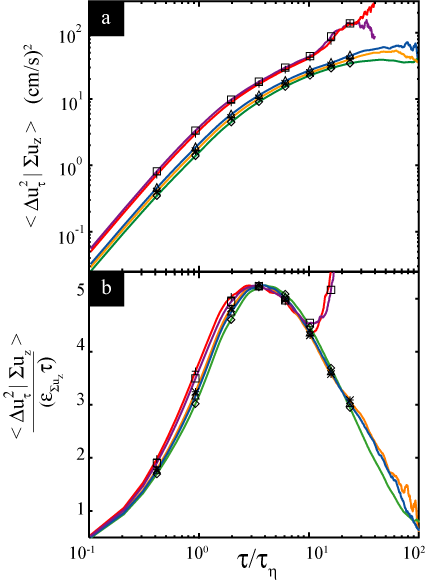}
\caption{\label{fig:Lsf2_zvel_center} (Color online) Second order
Lagrangian velocity structure function conditioned on instantaneous
velocity (vertical component) in the center of the tank.  a)
Uncompensated structure functions.  b) Individually compensated to
have the peak values match. Symbols represent the following
dimensionless vertical velocities, $\Sigma u_z/\sqrt{\langle u_z^2
\rangle}$: $+$ = 3.1 to 1.9, $\ast$ = 1.9 to 0.62, $\diamond$ = 0.62
to -0.62, $\triangle$ = -0.62 to -1.9 , $\Box$ = -1.9 to -3.1.}
 \end{center}
 \end{figure}
%

\subsubsection{Lagrangian structure functions conditioned on the large
scale velocity:  center region}
\label{sec:L_sf}

In much the same way we can evaluate the conditional Lagrangian
structure functions. Figure{~\ref{fig:Lsf2_zvel_center}(a)} shows
the second order Lagrangian structure function conditioned on the
vertical component of the large scale velocity, $\Sigma u_z$. Here we find
$\Sigma u_z$ by averaging the velocity of the particle at the two times
used to determine $\Delta u_\tau$. The conditional structure
functions for different large scale
velocity are different by a factor of about 2.5, and they remain nearly parallel throughout the entire time
range.

Figure{~\ref{fig:Lsf2_zvel_center}(b)} shows the second order
conditional Lagrangian structure function  compensated by
Eq.~\ref{2} where $\varepsilon$ is individually chosen so the maxima
of all of the conditioned structure functions coincide. This aids
comparison of the functional forms of the conditioned structure
functions. Again, the functional form is nearly identical for
different large scale velocities, indicating that the large scales
affect all time scales in the same way.  There may be a small trend
towards larger values of the compensated Lagrangian structure
functions at small times when the magnitude of the large scale
velocity is large.

It should be noted that there is a bias present in Lagrangian
measurements that is not present in Eulerian measurements.  A sample
of measured Lagrangian trajectories is biased towards low velocity
particles since the high velocity particles are more likely to have
left the measurement volume.  This bias becomes larger for larger
$\tau$. Berg \textit{et al.}.~\cite{Berg:2009} have studied this
bias and find that it can be quite large for typical experimental
conditions. We quantified this bias in our data by measuring the
Lagrangian structure functions using trajectories that remained
inside artificially restricted measurement volumes.  From a simple
extrapolation of the dependence on the size of the artificial
detection volume, we estimate that our experimental Lagrangian
structure functions underestimate the true value by 17\% for $\tau =
8 \tau_\eta$.  This is roughly consistent with the size of the error
we expect based on the critical time lag defined in
Ref.~\cite{Berg:2009}. Note that we have not performed the
compensation they recommend and we are roughly translating their
uncompensated results.  Because of this bias, we will focus
attention on $\tau < 10 \tau_\eta$. As we'll discuss in
section~\ref{sec:Lsf_cond_center}, the dependence of the conditioned
Lagrangian structure functions on the large scale velocity does not
seem to be significantly influenced by this bias.

%

\subsubsection{Eulerian structure functions conditioned on the large
scale velocity:  near grid region}

\begin{figure}[bth]
\begin{center}
\includegraphics[bb=0 0 640 750,scale=0.5]{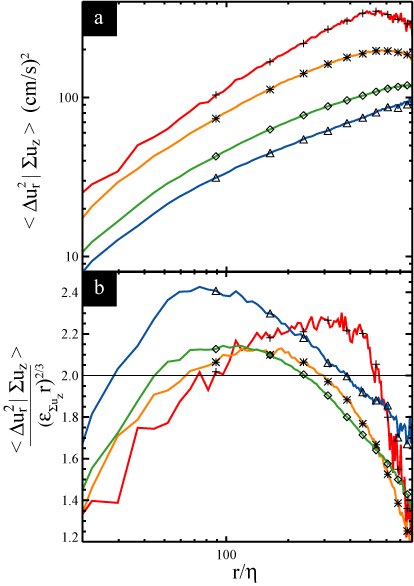}
\caption{\label{fig:sf2 zvel_bottom}(Color online) Second order
velocity structure function conditioned on particle pair vertical
velocity (Z direction) in the region near the bottom grid. The
condition with the largest downward velocity has been eliminated due
to lack of statistical convergence. Symbols represent the following
particle pair vertical velocities $\Sigma u_z/\sqrt{\langle u_z^2
\rangle}$: $+$ = 3.8 to 2.3, $\ast$ = 2.3 to 0.75, $\diamond$ = 0.75
to -0.75, $\triangle$ = -0.75 to -2.3, a) Uncompensated structure
functions b) Individually compensated by the energy dissipation rate
for each conditional data set.}
 \end{center}
 \end{figure}

By comparing separate regions of the tank we are able to explore the
effects of inhomogeneity on this conditional dependence.
Figure{~\ref{fig:sf2 zvel_bottom}} shows the Eulerian structure
functions, similar to Fig.{~\ref{fig:sf2_zvel_center}}, but with
data collected in the inhomogeneous region near the bottom grid
(NG). The separation between Eulerian structure function conditions
has doubled to approximately a factor of four. Note the different
ordering of the structure functions. The up-down symmetry is now
broken. Fluid traveling upwards ($*$ symbols) has a large structure
function while fluid traveling downward with the same magnitude of
vertical velocity ($\triangle$ symbols) has the lowest
value of the structure function. We interpret this as highly                
energetic fluid originating near the bottom grid and being
turbulently advected into the observation volume.  Similarly, fluid
carried down from the more quiescent region above the detection
volume has low energy and a smaller structure function.

Figure~\ref{fig:sf2 zvel_bottom}(b) shows the compensated Eulerian
structure functions, similar to figure~\ref{fig:sf2_zvel_center}(b),
but reveals a novel insight.  Stepping through the vertical velocity
bins is equivalent to stepping through the energy cascade.  Fluid
coming directly upward from the bottom grid (symbol $+$) carries
energy that was recently injected into the large scales.  As a
result, the compensated structure function for upward moving fluid
is biased towards the large scales. Fluid that has downward vertical
velocity (symbol $\triangle$) comes from the center region far away
from the grid. It has had more time to mature, and in this process
the energy is transported to smaller length scales. Conditional
structure functions appear to be an effective tool to evaluate
whether or not a turbulent flow is fully developed and has
established a stable cascade.

\begin{figure}[tbh]
\begin{center}
\includegraphics[bb=0 0 640 480,scale=0.5]{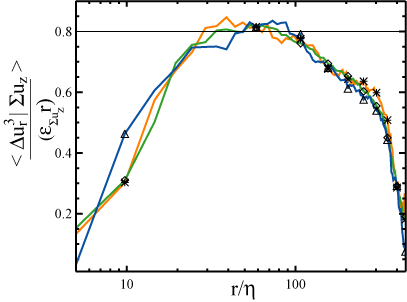}
\caption{\label{fig:sf3_zvel_center} (Color online) Third order
velocity structure function plots conditioned on particle pair
vertical velocity and individually compensated for each conditional
data set. Data is taken in the center region of the tank, and the
extreme vertical velocity plots have been eliminated due to lack of
statistical convergence. Symbols represent the following vertical
velocities $\Sigma u_z/\sqrt{\langle u_z^2 \rangle}$: $\ast$ = 2.5
to 0.84, $\diamond$ = 0.84 to to -0.84, $\triangle$ = -0.84 to
-2.5.}
 \end{center}
 \end{figure}

\subsubsection{Third order Eulerian structure functions conditioned on the large
scale velocity:  center region}

Figure~\ref{fig:sf3_zvel_center} shows the third order structure
function individually compensated and conditioned on $\Sigma u_z$ in the
center of the tank.  Convergence of third order statistics was more
difficult, so elimination of the two extreme conditions was
required. The third order structure function proves to be similar to
the second order in separation, symmetry, and collapse to a single
functional form. The energy dissipation rates found for the three
conditions are: $\varepsilon_\ast= 25.2 $cm$^2/$s$^3$,
$\varepsilon_\diamond = 21.7 $cm$^2/$s$^3$,$\varepsilon_\triangle =
28.7 $cm$^2/$s$^3$.




\subsection{A Powerful Method for Plotting Conditional Structure Functions}
\label{sec:Sreenigraphs}

\subsubsection{Eulerian structure functions conditioned on the large
scale velocity:  center region}

\begin{figure}[bth]
\begin{center}
\includegraphics[bb=0 0 640 480,scale=0.5]{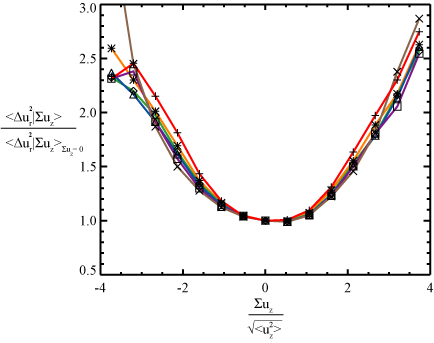}
\caption{\label{fig:sreenigraph_wo_sreeni_center} (Color Online)
Eulerian second order conditional structure function versus large
scale velocity. Data taken in the center region. Each curve
represent the following separation distances $r/\eta$: $+$ = 0 to
40, $\ast$ = 40 to 70, $\diamond$ = 70 to 110, $\triangle$ = 110 to
140, $\Box$ = 300 to 370, $\times$ = 370 to 440.}
 \end{center}
 \end{figure}

An alternative, and in many ways a more powerful, method of visualizing the same data is presented                    
in Fig.~\ref{fig:sreenigraph_wo_sreeni_center}.  Here we show the
second order Eulerian structure function conditioned on the vertical
component
of the large scale velocity (the same                   
data as Fig.{~\ref{fig:sf2_zvel_center}}).  However, the scaled
vertical pair velocity is plotted on the horizontal axis with
conditioned structure functions on the vertical axis.  When the
structure functions are scaled by their value at $\Sigma u_z = 0$,
we find very good collapse of the data.  The fact that these curves
for different $r/\eta$ collapse so well is a striking demonstration
that the large scales affect all length scales in the same way.  The
fact that the conditional structure functions vary by a factor of
2.5 demonstrates the strong dependence on the large scales.  Note
that for Gaussian random fields, the plot in
Fig.~\ref{fig:sreenigraph_wo_sreeni_center} would be flat, and a
nearly flat result is observed in DNS and grid
turbulence~\cite{sreenivasan:1998}.

 In Fig.~\ref{fig:sreenigraph_wo_sreeni_center}, it may be
expected that the structure function at the largest length scales
($\times$) are a function of the large scale velocity. We see here
that the dependence is a steep parabola.  What is now more clear
with this plotting method is the extent to which all the smaller
scales are also affected by the large scale velocity; in fact, all
length scales collapse nearly perfectly onto one parabola. The large
scale velocity affects all length scales in nearly the exact same
way, all the way down to the dissipative range.

 \begin{figure}[bth]
\begin{center}
\includegraphics[bb=0 0 640 480,scale=0.5]{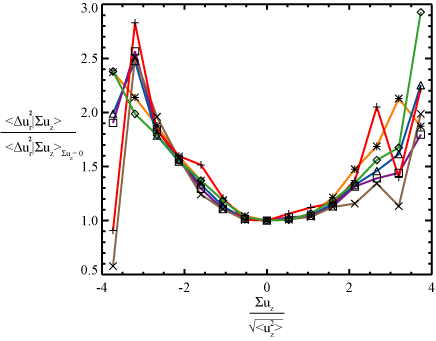}
\caption{\label{fig:sreenigraph center_5Hz} (Color Online) Eulerian
second order conditional structure function versus large scale
velocity. Data taken in the center region at higher grid frequency,
5Hz, resulting in higher Taylor Reynolds number ~380. Symbols
represent the following separation distances $r/\eta$: $+$ = 0 to
50, $\ast$ = 50 to 100, $\diamond$ = 100 to 150, $\triangle$ = 150
to 200, $\Box$ = 310 to 420, $\times$ = 420 to 520.}
 \end{center}
 \end{figure}

\subsubsection{Eulerian structure functions conditioned on the large
scale velocity:  higher Reynolds number}

Figure \ref{fig:sreenigraph center_5Hz} shows the effect of
increasing the Reynolds number.  This data is at the center of the
tank with the grids oscillating at 5 Hz which increases $R_\lambda$
to 380.  The collapse of the structure function remains.  The
curvature in this figure is not significantly different from the
lower Reynolds number data in
Fig.~\ref{fig:sreenigraph_wo_sreeni_center} indicating that if there
is a Reynolds number dependence it is weak.

 \begin{figure}[bth]
\begin{center}
\includegraphics[bb=0 0 640 480,scale=0.5]{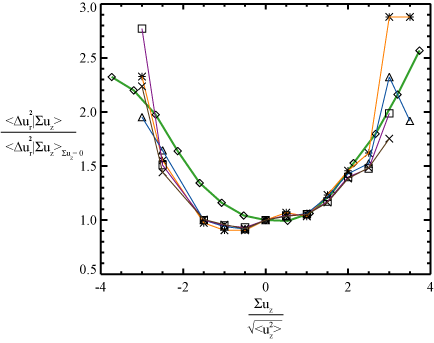}
\caption{\label{fig:sreenigraph sreeni_data_center} (Color Online)
Eulerian second order conditional structure function versus large
scale velocity. The thin plots are from atmospheric boundary layer
data ~\cite{sreenivasan:1998} $r/\eta$: $\ast$ $\thicksim$ 100,
$\triangle$ $\thicksim$ 400, $\Box$ $\thicksim$ 1000, $\times$
$\thicksim$ 1250. The thick line is from
fig.~\ref{fig:sreenigraph_wo_sreeni_center}, which has been overlaid
for comparison, $r/\eta$: $\diamond$ = 70 to 110.}
 \end{center}
 \end{figure}

Figure~\ref{fig:sreenigraph sreeni_data_center} shows a comparison
of our data with data taken in the atmospheric boundary
layer~\cite{sreenivasan:1998} with $R_\lambda
> 10^4$.  Atmospheric boundary layer turbulence
shows a similar collapse of conditional structure functions at all
length scales.  The curvature is also similar in both data sets,
indicating that the dependence on the large scales is similar even
at these very large Reynolds numbers.

 \begin{figure}[bth]
\begin{center}
\includegraphics[bb=0 0 640 480,scale=0.5]{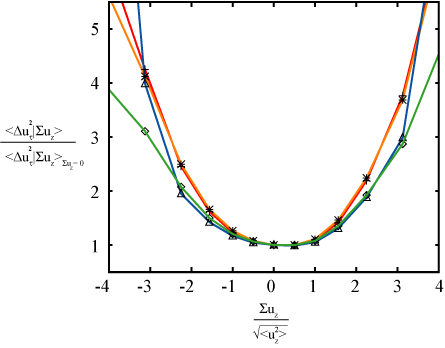}
\caption{\label{fig:Lsreenigraph_wo_sreeni_center} (Color Online)
Lagrangian second order conditional structure function versus large
scale vertical velocity. Data taken in the center region. Symbols
represent the following
 $\tau/\tau_\eta$: $+$ = 0.42 , $\ast$ = 1.3,
$\diamond$ = 3.5, $\triangle$ = 10.}
 \end{center}
 \end{figure}

 \subsubsection{Lagrangian structure functions conditioned on the large
scale velocity:  center region}
\label{sec:Lsf_cond_center}

Figure~\ref{fig:Lsreenigraph_wo_sreeni_center} shows the Lagrangian
structure functions plotted versus the large scale velocity,
comparable to the Eulerian data shown in
Fig.~\ref{fig:sreenigraph_wo_sreeni_center}.  The parabolic shape
remains, but the curvature is greater for all Lagrangian time scales
than it is in the Eulerian data.  All time scales are affected by
the large scale velocity.  To determine the effect of measurement volume bias,
we have done this analysis for artificially restricted measurement volumes.  By decreasing
the volume by a factor of 2, we observe the large $\tau$ curves to shift by approximately the
deviations between the curves.
We conclude that the bias does not have a significant effect on the
conditional dependence shown in Fig.~\ref{fig:Lsreenigraph_wo_sreeni_center} for the
time differences presented  ($\tau \le 10 \tau_\eta$).

\begin{figure}[bth]
\begin{center}
\includegraphics[bb=0 0 640 480,scale=0.5]{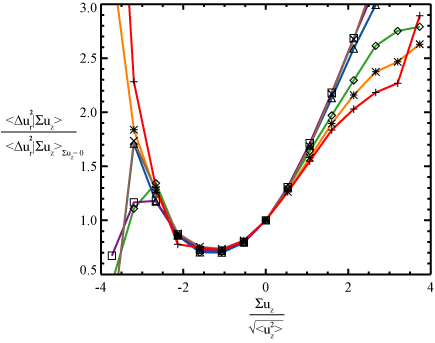}
\caption{\label{fig:sreenigraph bottom 3Hz} (Color Online) Eulerian
second order conditional structure function versus large scale
velocity. Data taken in the near grid region of the tank. The
structure function is heavily influenced by the bottom grid which
has skewed the symmetry of the plot minima in the negative
direction. Symbols represent the following non-dimensional
separation distances $r/\eta$: $+$ = 0 to 50, $\ast$ = 50 to 110,
$\diamond$ = 110 to 160, $\triangle$ = 270 to 320, $\Box$ = 330 to
450, $\times$ = 450 to 560.}
 \end{center}
 \end{figure}

  \subsubsection{Eulerian structure functions conditioned on the large
scale velocity:  near grid region}

Figure~\ref{fig:sreenigraph bottom 3Hz} shows conditioned Eulerian
structure function similar to
figure~\ref{fig:sreenigraph_wo_sreeni_center}, but measured in the
inhomogeneous region near the grid (NG).  The structure functions
here are strikingly different than in the center.  The minimum is
shifted by more than one standard deviation to the left. The
inhomogeneity breaks the up-down symmetry so that fluid coming
directly up from the bottom grid is markedly different then fluid
coming down from the more quiescent region above (analogous to the
$*$ and $\triangle$ separation in Fig.~\ref{fig:sf2 zvel_bottom}).
 It follows that fluid with an upward
velocity has higher energy than fluid with the same velocity
magnitude in the downward direction. The atmospheric boundary layer
data in Fig.~\ref{fig:sreenigraph sreeni_data_center} also shows
this effect with a minimum at $\Sigma u_z/\sqrt{\langle u_z^2
\rangle}=-0.5$, presumably as a result of weaker inhomogeneity. Also
notable is that the collapse of plots for various $r$ values is not
as complete as in the central region.  This is consistent with
Fig.~\ref{fig:sf2 zvel_bottom}(b) which shows that the conditional
structure functions have somewhat different $r$ dependence.

 \begin{figure}[bth]
\begin{center}
\includegraphics[bb=0 0 640 480,scale=0.5]{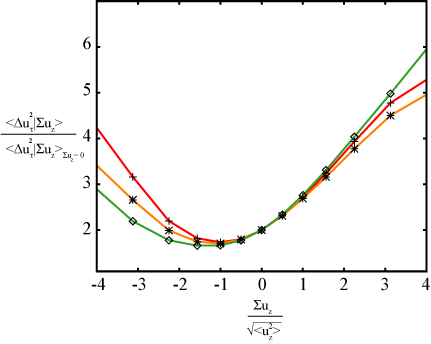}
\caption{\label{fig:Lsreenigraph_wo_sreeni_bottom} (Color Online)
Lagrangian second order conditional structure function versus large
scale vertical velocity. Data taken in the near grid region of the
tank. Symbols represent the following
 $\tau/\tau_\eta$: $+$ = 0.94, $\ast$ = 2.8,
$\diamond$ = 8.0.}
 \end{center}
 \end{figure}

  \subsubsection{Lagrangian structure functions conditioned on the large
scale velocity:  near grid region}

Figure~\ref{fig:Lsreenigraph_wo_sreeni_bottom} shows a Lagrangian
structure function taken in the near grid region, similar to the
Eulerian data in Fig.~\ref{fig:sreenigraph bottom 3Hz}.  The minimum
is shifted to the left here also as a result of the inhomogeneity in
this region of the flow.  The conditional dependence on the large
scale velocity is again somewhat larger than in the Eulerian case,
and the collapse at different time scales is not as complete.

\begin{figure}[bth]
\begin{center}
\includegraphics[bb=0 0 640 480,scale=0.5]{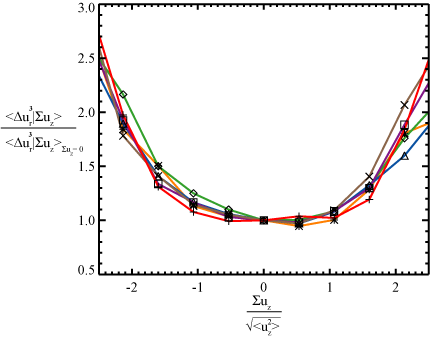}
\caption{\label{fig:sf3_rbins8_zvelbins16_center 3Hz} (Color Online)
Eulerian third order conditional structure function versus large
scale vertical velocity in the center region. Symbols represent the
following non-dimensional separation distances $r/\eta$: $+$ = 0 to
40, $\ast$ = 40 to 70, $\diamond$ = 70 to 110, $\triangle$ = 110 to
140, $\Box$ = 220 to 300, $\times$ = 300 to 370.}
 \end{center}
 \end{figure}

   \subsubsection{Third order Eulerian structure functions conditioned on the large
scale velocity:  center region}

The third order Eulerian velocity structure function plotted versus
the large scale vertical velocity is shown in
Fig.~\ref{fig:sf3_rbins8_zvelbins16_center 3Hz} using data from the
center of the tank.   Statistical convergence is weaker than the
second order which limits the large scale velocity range available
for analysis.  The collapse seems similar to the second order case
shown in Fig.~\ref{fig:sreenigraph_wo_sreeni_center}, although the
measurement uncertainties are larger here. The curvature seems to be
slightly larger for the third order than for the second order case.

\begin{figure}[bth]
\begin{center}
\includegraphics[bb=0 0 640 480,scale=0.5]{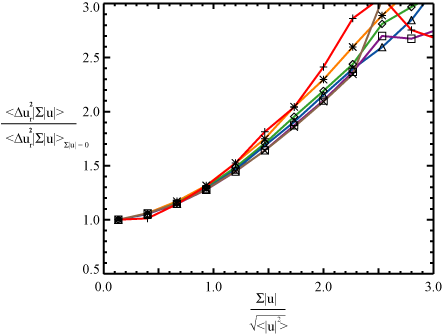}
\caption{\label{fig:sf2_rbins8_zvelbins15_RMS_center 3Hz} (Color
Online) Eulerian second order conditional structure function versus
magnitude of the velocity pair in the center region. Symbols
represent the following non-dimensional separation distances
$r/\eta$: $+$ = 0 to 40, $\ast$ = 40 to 70, $\diamond$ = 70 to 110,
$\triangle$ = 110 to 140, $\Box$ = 300 to 370, $\times$ = 370 to
440.}
 \end{center}
 \end{figure}

   \subsubsection{Second order Eulerian structure functions conditioned on the
velocity magnitude:  center region}

The second order Eulerian structure function plotted versus the
magnitude of the pair velocity is shown in
Fig.~\ref{fig:sf2_rbins8_zvelbins15_RMS_center 3Hz} using data from
the center observation volume.  The magnitude of the pair velocity
is also a useful indicator of large scale activity. It has no
preferred direction, and it is a more direct indicator of the
instantaneous local energy. A similar dependence remains as in
Fig.~\ref{fig:sreenigraph_wo_sreeni_center}, the collapse seems
similar, and the curvature is significantly larger.

%

\subsection{Discussion}
\label{sec:Discussion}

We have provided a comprehensive set of measurements that shows
signatures of the current state of the large scales on the inertial
range and small scales in turbulence.   Here we wish to discuss
factors that might be responsible for
the dependence of structure functions on the instantaneous large
scale velocity.  First we will address a possible concern that the conditional
dependence may be a kinematic correlation.  Then we will discuss possible
properties of the large scales that could be
important including Reynolds number, anisotropy, mean shear,
inhomogeneity, and large scale intermittency.    We do not claim that this list is exhaustive, but
it seems that identifying the major contributors will be valuable.

A reasonable suspicion might be that the observed dependence is a kinematic correlation, meaning that particle
pairs with large velocity may also have a large velocity
difference simply because the same measurements are used in both
cases.  Hosokawa~\cite{Hosokawa:2007} identified that Kolmogorov's 4/5ths law requires
that velocity sums and differences be correlated so that
\begin{equation}
\langle u_+^2 \Delta u_- \rangle = \frac{\epsilon r}{30}
\end{equation}
where $u_-$ is half the longitudinal velocity difference and $u_+$
is half the sum.  (For comparison, we have used $\Delta u_r=2u_-$
and below $\Sigma u_\parallel=u_+.$) Khomyansky~\cite{Tsinober:2008}
et al provide an experimental confirmation of this and in a more
recent paper~\cite{Tsinober:preprint} present a list of kinematic
relations. However, several lines of evidence indicate that
kinematic correlation does not account for the majority of the
dependence we observe.

First, note that two independent random samples with identical Gaussian distributions have a difference that
is uncorrelated with the sum, so that the conditional dependence seen in Fig~\ref{fig:sreenigraph_wo_sreeni_center}
would be flat.  This remains true for velocity differences and sums from Gaussian
random fields.  Both of these results can be obtained by considering the joint pdf
of the two samples, and then rotating 45 degrees to the coordinate system of sums and
differences.  Because the samples are interchangeable, the sum and difference axes have
to be principle axes of the joint gaussian pdf, and the conditional variance of the difference
is independent of the sum.

Of course, turbulent velocities are not joint Gaussian.   However,
from the kinematic relations in the literature we have not been able
to derive predictions for the conditional structure functions that
we consider or for the correlation  $\langle (\Sigma u)^2 (\Delta
u)^2 \rangle$ that would capture the main conditional dependence we
see.

 To make an experimental estimate of the effect of kinematic correlations, we       
conditioned the velocity differences on several other quantities.
For each particle pair, we calculated the longitudinal and
transverse components of the average velocity of the particles,
denoted $\Sigma u_\parallel$ and $\Sigma u_\perp$ respectively. We
then conditioned the longitudinal structure functions on the
longitudinal and transverse pair velocities instead of conditioning
on the pair vertical velocity. The idea here is that while
conditioning the longitudinal structure functions on the
longitudinal component ($\Sigma u_\parallel$) could have a kinematic
correlation, conditioning on the transverse component ($\Sigma
u_\perp$) should have no kinematic correlation. We found that
conditioning on ($\Sigma u_\parallel$) had a roughly 30\% larger
effect on the structure functions than conditioning on ($\Sigma
u_\perp$). Conditioning on $\Sigma u_z$ should have less kinematic
correlation than conditioning on $\Sigma u_\parallel$. So more than
70\% of the effect remains unexplained by kinematic correlation. We
conclude that while kinematic correlation may possibly make a
significant contribution to the conditional dependence, the majority
of the effect comes from the large scales.

An immediate concern when discussing large scale effects is if the oscillating grid flow has a Reynolds                 
number insufficient for adequate scale separation, and it is this
which leads to contamination of the small scale statics by the large
scales. Evidence points to large scale dependence not being caused
by limited Reynolds number. The comparison in
Fig.~\ref{fig:sreenigraph sreeni_data_center} shows that atmospheric
boundary layer data~\cite{sreenivasan:1998} with very large Reynolds
number ($R_\lambda > 10^{4}$) has nearly the same dependence on the
large scales as our flow. Increasing the Reynolds number in our flow
makes very little difference.  Additionally, all length scales
collapse to nearly the same functional form indicating that limited
separation of scales is not the primary factor.  Taken together,
these lead us to the conclusion that merely high Reynolds number
alone is not enough to create small scales that are statistically
independent of the large scales.

Our flow is somewhat anisotropic.  The ratio of vertical to
horizontal velocity standard deviations is 1.5:1 in the center.  The
effects of large scale anisotropy on the small scales has been
studied extensively~\cite{biferale:1999}, but it is not a major
factor in conditional dependence studied here. We have analyzed our
data by averaging over particle pairs with all orientations, so when
the structure functions are conditioned on a quantity with no
preferred direction like the velocity magnitude (Figure
~\ref{fig:sf2_rbins8_zvelbins15_RMS_center 3Hz}) there should be
very little contribution from anisotropy.   In fact, we find that
the conditional dependence on velocity magnitude is even stronger
than the dependence on the vertical velocity component.  We also
observe the conditional dependence remains when conditioned on other
quantities without preferred directions like $\Sigma u_\parallel$,
and $\Sigma u_\perp$.  We conclude that anisotropy of the large
scales is not a significant cause of the conditional dependence we
observe.


Sreenivasan and Dhruva~\cite{sreenivasan:1998} attribute the strong
conditional dependence of the Eulerian structure functions on the
large scale velocity to shear in the atmospheric boundary layer. In
making this argument, they show an important piece of information in
their figure 6 which shows conditioned structure functions in
homogeneous turbulence from both DNS and wind tunnel grid
turbulence.  The conditional statistics in these homogeneous and
isotropic flows show no apparent dependence on the large scale       
velocity.  However, we conclude that shear is not the fundamental
property responsible in our flow since the oscillating grid flow has
a much lower shear but produces much the same dependence on the
large scale velocity. The mean velocity gradient normalized with the
eddy turnover time is 1.2 in the center of our flow and we estimate
it is in the range of 5 or greater for their atmospheric boundary
layer data.  There must be some other properties that exist in the
shear flow, but also are important in our flow with small shear.

Our data clearly shows the role that inhomogeneity plays in the
observed large scale dependence.  Our Eulerian and Lagrangian data
near the grid in Figs.~\ref{fig:sreenigraph bottom 3Hz} and
\ref{fig:Lsreenigraph_wo_sreeni_bottom} show that the structure
functions depend greatly on the origin of the fluid being swept into
the observation volume. Fluid coming from energetic regions of the
tank have larger structure functions than fluid coming from more
quiescent regions.  Inhomogeneity is directly responsible for the
shift of the minimum in Fig.~\ref{fig:sreenigraph bottom 3Hz} away
from zero vertical velocity.  In the center of the tank
(Fig.~\ref{fig:sreenigraph_wo_sreeni_center}), the inhomogeneity is
much smaller, but it could be responsible for part of the curvature
since both fluid coming downward and fluid coming upward would be
coming from more energetic regions symmetrically.

However, inhomogeneity alone does not account for all of the large
scale dependence observed.  There is also a significant contribution
from large scale intermittency, and it is possible that this is the
dominant contribution in the center of the tank.  Large scale
intermittency has been difficult to quantify.  It can be defined as
any temporal fluctuations in the large scales that occur on
timescales longer than the eddy turnover time, $L/u$.

Fernando and DeSilva \cite{Fernando:1993} show large scale
intermittency can exist in an oscillating grid flow depending on
boundary conditions.  We have observed clear signatures of large
scale intermittency in our flow.  Although we use their recommended
boundary conditions, the velocity distribution in the center of the
flow is slightly bimodal indicative of switching between two flow
states. This effect is more prominent in preliminary data we took
for grid spacings of 66 cm and 100 cm than it is in the data for
56.2cm presented in this paper.

Our measurements show a dependence of the conditional structure
functions on the large scale velocity that can not be fully
attributed to inhomogeneity, and large scale intermittency appears
to be the most likely cause. The clearest evidence for this comes
from conditioning the structure functions on the horizontal
components of the large scale velocity, $\Sigma u_x$ and $\Sigma u_y$ instead of
on the vertical component, $\Sigma u_z$.
 The horizontal midplane ($x$ and $y$ directions) is much more                          
homogeneous than the vertical axis (z direction). Yet, the structure
functions conditioned on $\Sigma u_y$ or $\Sigma u_x$ show a large scale
dependence that is only moderately smaller than for the $\Sigma u_z$
condition (85\% and 72\% of the dependence seen in $\Sigma u_z$). If the
inhomogeneous direction shows similar conditional dependence on the
large scales as the homogeneous directions show, then it seems that
a large part of the conditional dependence must come from
fluctuations in the large scales, and not directly from
inhomogeneity.
 Praskovsky \textit{et al.}~\cite{Praskovsky:1993}
attribute large scale intermittency as a crucial component of the
large scale dependence they observe. More work is needed to isolate
the effects of large scale intermittency on small scale statistics
in turbulence.

We have largely ignored considerations of power law scaling which
has been a focus of much of the previous work on this subject.
Because of the relatively low Reynolds number of our experiment, we
can  not make sensitive tests of scaling.  However, our data
provides a plausible picture about how the large scales should
affect power law scaling.  If the data in
Fig.~\ref{fig:sreenigraph_wo_sreeni_center} collapses to a single
curve, then the dependence of the conditional structure functions on
$r$ and $u_z$ are separable and the large scale dependence will have
no effect on the scaling exponents of unconditional structure
functions. When this type of plot does not collapse as in
Figs.~\ref{fig:sreenigraph bottom 3Hz} and
~\ref{fig:Lsreenigraph_wo_sreeni_bottom}, then the power law scaling
will be affected by the large scales.

\section{Conclusions}           
We study a flow between oscillating grids with 3D particle tracking
and a novel real-time image compression system in order to quantify
the effects of various properties of the non-universal large scales
on the inertial range and small scales.

We measured the mean and variance of the velocity as a function of
distance from the grids.  The oscillating grid motion has produced a
weak mean flow as well as a region near the grid with high velocity
variance that falls off quickly to a very homogeneous, lower
velocity variance, region in the center. This profile has been key
in the determination of the role of inhomogeneity.

Conditional statistics were employed in order to measure the large
scale effects. Second order Eulerian velocity structure
functions were conditioned on the phase of the grid, an obvious
source for periodic large scale energy input. Results show little
dependence of the structure functions in the center region and
surprisingly little even near the grid.

Eulerian and Lagrangian structure functions were also conditioned on
the instantaneous large scale velocity. A large dependence was found
in the center, with the Eulerian structure function increasing by a
factor of 2 or more when the large scale velocity is large.  The
dependence of the Lagrangian structure functions is somewhat larger.
Conditioned structure functions show that in the center of the tank,
all length scales are affected in approximately the same way. The
region near the grid was also analyzed and compared with the region
in the center.  Near the grid, we found a much stronger dependence
on the instantaneous large scale velocity for both the Eulerian and
Lagrangian structure functions than we found in the center. Near the
grid, there are clear signatures of the effects of large scale
inhomogeneity on the small scales.  Fluid coming up from the
energetic region nearer the grid has large structure functions,
while fluid coming down from the quiescent region in the center has
much smaller structure functions.  The functional form of the
conditional structure functions are also different indicating the
different histories of the different fluid.  These measurements
provide a clear picture of the way inhomogeneity affects the small
scales of turbulence.

Plotting the conditional structure functions versus the large scale
velocity provides a powerful method for visualizing the effects of
the large scales on all length scales in turbulent flows. We
recommend these plots as an effective way to compare the effects
of the large scales in different experiments.  This has been done for grid turbulence and     
homogeneous, isotropic DNS~\cite{sreenivasan:1998} which show almost
no dependence of the structure functions on large scale velocity.
Our oscillating grid flow and high Reynolds number atmospheric
boundary layer turbulence~\cite{sreenivasan:1998} show very similar
dependence.  Comparison of conditional structure functions in other
flows has the potential to clarify the effects of the large scales
on small scale turbulence and to guide the search for universal
properties of turbulent flows.

%
%
%

\section{\label{sec:acknowledgements}Acknowledgements}
This work was supported by Wesleyan University, the Alfred P. Sloan
foundation, and NSF grant DMR-0547712.  We thank Rachel Brown,
Emmalee Riegler and Tom Glomann for assistance with the experiment.
We benefitted from discussions with Nick Ouellette, Haitao Xu,
Eberhard Bodenschatz, Zellman Warhaft, Laurent Mydlarski, and Mark
Nelkin.


\end{document}